# Performance of a Chamber for Studying the Liquid Xenon Response to Nuclear Recoils

V. Chepel, F. Neves, V. Solovov, A. Pereira, M. I. Lopes, J. Pinto da Cunha, P. Mendes, A. Lindote, C.P. Silva, R. Ferreira Marques and A. J.P.L. Policarpo

*Abstract* — The design and performance of a 1.2 liter liquid xenon chamber equipped with 7 two-inch photomultiplier tubes, with the purpose of studying the scintillation response of xenon to γ-rays and neutrons, is described. Measurements with γ-rays indicate a high VUV light collection efficiency resulting in ~5.5 photoelectrons per 1 keV of deposited energy. The energy resolution (FWHM) is 18% and 22%, for 122 keV and 511 keV γ-rays, respectively. An algorithm for the reconstruction of the scintillation coordinates in (*x*,*y*) plane was developed and tested. The position resolution is estimated to be σ=6.9 mm for 122 keV γ-rays.

## I. INTRODUCTION

LIQUID xenon detectors are increasingly considered as a powerful instrument for a direct search of weakly interactive massive particles (WIMP) which might constitute a significant fraction of the non-luminous matter in the Universe [1]. Due to the proximity of the mass of the xenon nucleus (A=130) to some estimates of the WIMP mass, the energy transfer from a WIMP to xenon nuclei in elastic collisions is maximized. As the expected energy distribution of nuclear recoils decreases rapidly with energy, the detection threshold must be as low as possible. At present, several experiments are aiming at achieving sensitivity to nuclear recoils in the energy range from a few keV to some tens of keV. Thus, detailed knowledge of the response of liquid xenon to low energy nuclear recoils is of the utmost importance.

Both scintillation efficiency and the decay time of liquid xenon are known to depend on the ionization density along the particle track. The determination of the light yield per unit of deposited energy for nuclear recoils with respect to that for electrons of the same energy (frequently called quenching factor) is important for accessing the energy spectrum of detected recoils, given that the detector calibration is usually (and more conveniently) performed with γ-rays. Moreover, the differences in the scintillation decay times from nuclear recoils and from electrons allow these signals to be distinguished and the background from γ and β rays rejected.

Given its high atomic number, high density, high scintillation light yield and fast decay time, liquid xenon is also known to be a very good medium for efficient detection of γ-rays. These properties are intended to be exploited in a 4π γ-ray scintillation calorimeter for measurements of neutron capture cross sections in a wide range of neutron energies, whose feasibility study is in the program of the n-TOF Collaboration at CERN [2]. Such a calorimeter should detect γ-ray cascades following neutron capture in a target nucleus with 100% efficiency, good energy resolution and high counting rate. It has to be operated at a neutron beam under a significant background of scattered neutrons. Again, the separation of the signals due to γ-rays from those due to neutrons is very important.

In the present paper we describe the design of a liquid xenon scintillation chamber which was built to study the scintillation efficiency and decay time due to nuclear recoils and γ-rays. First results are also presented. Although the chamber is presently operated without electric field, its design allows future studies under an applied electric field.

## II. CHAMBER DESIGN

The chamber, the thermal insulation, as well as the liquefaction and cooling techniques, were designed having in mind the need to minimize the amount of material around the liquid xenon volume, in view of its operation in a neutron beam.

### A. Chamber

A simplified drawing of the chamber is shown in Fig. 1. The liquid xenon chamber is a stainless steel cylindrical vessel (∅200 mm, 278 mm high) with a thickness of 0.4 mm in the lateral wall and 6 mm at the bottom. This vessel is suspended inside another cylinder kept under vacuum, for thermal insulation. The outer cylinder is made of aluminium, 0.8 mm thick, reinforced with ~3 mm of carbon fiber epoxy glued on the air side.

The active volume of liquid xenon (about 1.2 liter) is defined by PTFE reflectors arranged to form a cylinder, ∅163 mm and 55 mm high. The bottom of this cylinder is made from a 4 mm thick PTFE sheet, whereas the lateral surface and the top disk (with 7 openings for the PMTs) have a

Manuscript received November 13, 2004. This work was supported in part by the European Commission under Contract No.FIKW-CT-2000-00107 and by Fundação para a Ciência e Tecnologia, Portugal, under project POCTI/FNU/43729/2002. F. Neves, V. Solovov and A. Lindote were supported by the FCT fellowships SFRH/BD/3066/2000, SFRH/BPD/14517/2003 and SFRH/BD/12843/2003, respectively.

The authors are with LIP-Coimbra and CFRM of the Department of Physics of the University of Coimbra, 3004-516 Coimbra, Portugal (telephone: (+351)-239-410626, e-mail: neves@lipc.fis.uc.pt - F. Neves).

thickness of 1 mm. The active volume is viewed by an array of 7 photomultipliers Hamamatsu R2154 in direct contact with the liquid xenon, their entrance windows immersed in order to maximize the light collection. The photomultipliers have 2−inch quartz window and bialkaline photocathode. Metal fingers are deposited under the photocathode to reduce its resistivity at low temperature. The PMT photocathodes are kept at ground potential. The quantum efficiency at the wavelength of 175 nm varies from tube to tube, between 21% and 25% at room temperature, as quoted by the manufacturer.

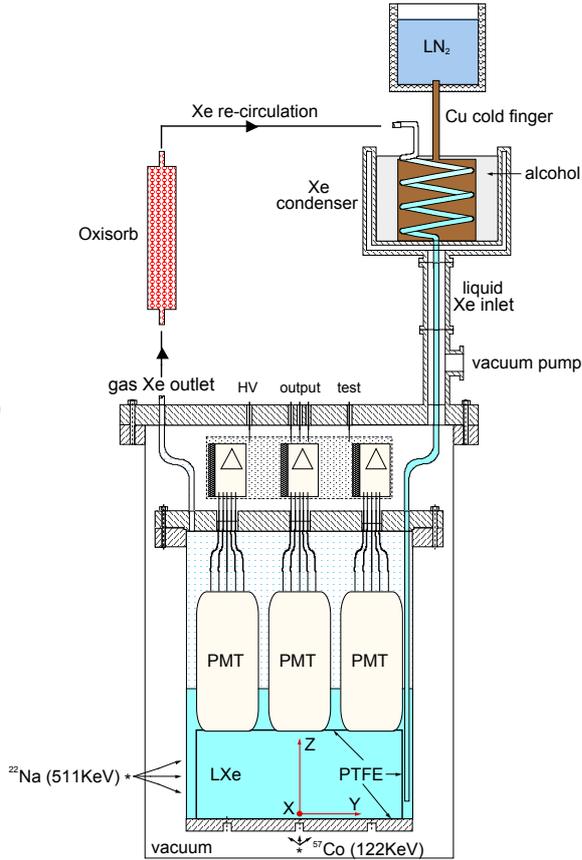

Figure 1. Schematic drawing of the chamber (to scale) and xenon condensation and re-circulation circuit (not to scale).

Xenon is introduced into the chamber through an inlet stainless steel tube, which passes in vacuum through the aluminium flange of the outer chamber. The condenser consists of a helix gas tube in good thermal contact with a massive copper block connected by a thick copper rod to a liquid nitrogen container. The condenser was filled with alcohol to improve the heat transfer and ensure a more uniform temperature along the helix.

### B. Xenon Purification, Liquefaction and Re-circulation

High purity commercial xenon gas was purified in a portable purification system by passing the gas through an Oxisorb column several times prior to condensing it into the chamber. The chamber and all connections to the purification system are primarily pumped down to ~$10^{-6}$ mbar under heating during several days. The connecting tubes are heated to a temperature of about 200 ºC, though the chamber is kept at 70 ºC, the maximum storage temperature quoted for these PMTs. Afterwards, xenon is introduced into the hot chamber. For the next several days, xenon gas is kept circulating through both the chamber and the Oxisorb column, at room temperature.

The chamber is cooled down by liquid xenon, which is condensed in the helix condenser (see Fig. 1) and drops into the chamber under gravity. The evaporated xenon gas is returned into the condenser through the chamber outlet, in a closed cycle. Once the chamber bottom reaches liquid xenon temperature, about −90 ºC at vapor pressure of 2.5 bar, the condensing of gas is started.

The liquid level is monitored during the condensation process by observing signals from the photomultipliers when a $^{57}$Co γ-ray source (122 keV) is placed below the chamber bottom, under its center. A sharp increase of the amplitude, by a factor of ~2, is observed when the liquid level is high enough to cover the entrance window of the respective photomultiplier. Condensing is continued for some time, allowing the liquid surface to raise well above the photomultiplier windows.

The chamber is operated at the xenon vapor pressure of 2.5 bar. During operation, the temperature of the helix condenser is always about −100 ºC. The temperature of the chamber is maintained stable by the xenon dripping in. The gas could be circulated either directly, or through the purification system, so that it can be permanently purified before returning to the condenser and chamber.

### III. FRONT END AND DATA ACQUISITION

Each of the 7 photomultiplier tubes is connected to a low noise charge-sensitive preamplifier followed by a shaping amplifier, both specially developed for this experiment [3], as shown in Fig. 2. The preamplifiers have an input noise of ~1 fC, 5 ns risetime and a falltime of 500 ns. The shaping amplifiers have an active CR-(RC)$^2$ filter with a characteristic time of 250 ns and an adjustable output gain. The signal from each preamplifier is also inverted, fed into a fast amplifier and then sent to a constant fraction discriminator. The outputs of these discriminators are sent to a logic sum unit to generate the trigger, by a coincidence of 2 or more signals. The trigger is used to generate both the GATE for the peak-ADCs and the START for the TDCs. The STOP signals for the TDCs are provided by the discriminators, after a suitable delay.

After integration by the charge-sensitive preamplifiers and filtering by the shaping amplifiers the signals from each PMT are sent to a multichannel peak-ADC unit. The timing signal of each photomultiplier is fed into a multichannel TDC unit. The ADC and TDC data are acquired by a computer and stored for further analysis. We refer to [3] for more details.

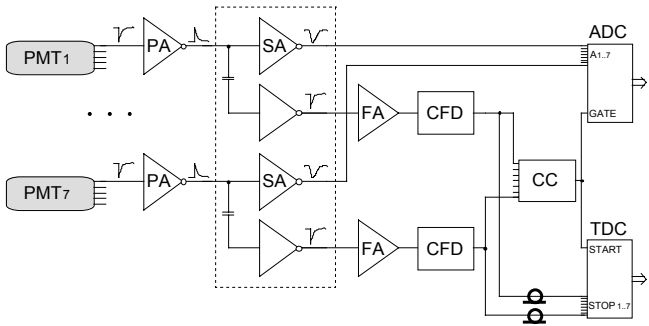

Figure 2. Schematic drawing of the front end and data acquisition system. PA –charge sensitive preamplifier, SA – shaping amplifier, FA – fast amplifier, CFD –constant fraction discriminator, CC – coincidence unit, ADC – amplitude to digital converter, TDC – time to digital converter.

## IV. RESULTS WITH GAMMA-RAYS

### A. Setup and Calibration

Measurements with γ-rays were performed using a $^{57}$Co (122 keV plus 136 keV in a 10:1 ratio) and $^{22}$Na (511 keV plus 1.2 MeV) sources. The $^{57}$Co source was placed under the chamber, inside the vacuum cryostat (see Fig. 1). The source can be moved from the outside and placed either at a fixed position under any of 7 photomultipliers or enclosed in a small lead container. Seven 1 mm thick and ⌀8 mm windows have been machined in the chamber bottom, below each photomultiplier (see Fig. 1). Additionally, a lead mask has been placed below the bottom of the chamber with ⌀8 mm holes aligned with the windows referred above.

For the 511 keV γ-rays, the chamber was irradiated laterally from a distance of about 40 cm. Pairs of 511 keV photons were selected by means of coincidences with a cylindrical ⌀2×2 inch crystal of $BaF_2$.

Each acquisition channel was calibrated with respect to the PMT gain at the operating temperature with a short light pulse (<300 ns) from a green LED by acquiring the single electron spectra from the respective PMT. The LED was placed outside the liquid xenon chamber in the vacuum shell close to the electrical feedthroughs, so that the light emitted by the diode could penetrate into the chamber through the ceramic insulators. The calibration was periodically checked during the measurements.

### B. Light Collection Efficiency and Energy Resolution

The pulse height spectrum measured with 122 keV γ-rays for the source placed below the central window is shown in Fig. 3 (a). The spectrum in Fig. 3 (b) was obtained with 511 keV γ-rays at the conditions referred above. All signals from all photomultiplier tubes were recorded, provided that a trigger occurred. Offline, after calibration, the PMT amplitudes are converted to number of photoelectrons and added with equal weights. An energy resolution (FWHM) of about 18% and 22% was obtained for 122 keV and 511 keV, respectively.

It is important to point out that the light collection is quite different for the two γ-ray energies. For 122 keV γ-rays the attenuation length in liquid xenon is about 3 mm and photo-absorption dominates. Thus, the majority of the γ-rays that reach the active volume are absorbed in a very thin layer of the liquid, compared to the height of the chamber. For 511 keV γ-rays, though, the attenuation length is about 3.5 cm and photoelectric absorption is only 21%. Moreover, the coincidence cone used in these measurements was rather large causing the γ-ray interactions to be widely distributed throughout the active volume. Therefore, the non-uniformity of the VUV light collection over the chamber volume is higher in the case of the more energetic γ-rays (see Fig. 4). This fact may explain the worse energy resolution for 511 keV than for 122 keV, contrary to what could be expected, taking into account only the number of scintillation photons produced in liquid xenon.

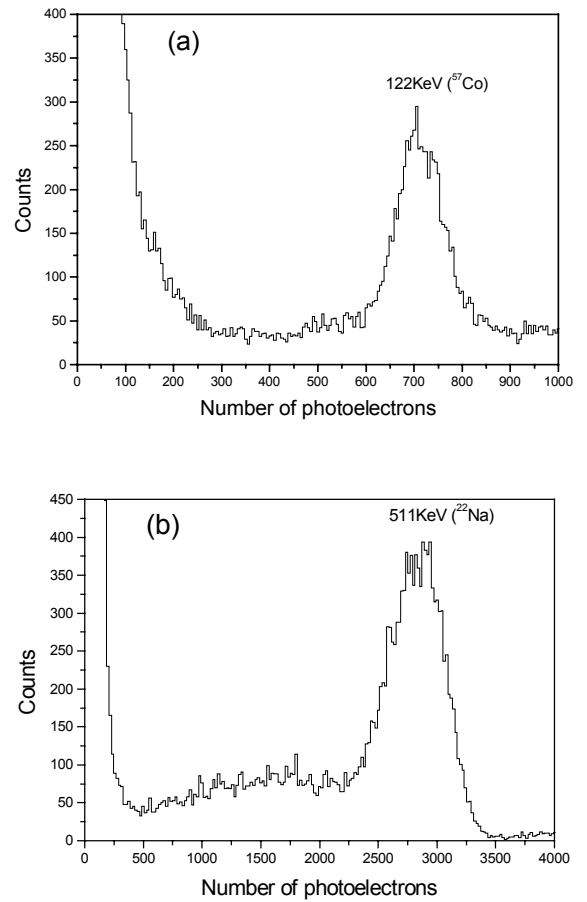

Figure 3. Pulse height spectra due to 122 keV (a) and 511 keV (b) γ-rays obtained by offline summing of the amplitudes of the 7 photomultipliers.

From these measurements we arrive at an overall conversion efficiency of ~5.5 photoelectrons per 1 keV of deposited energy. This corresponds to a collection of about half the emitted VUV photons, assuming that the mean energy required to produce one photon is $W_s = 20$ eV [4]-[6] and the quantum efficiency is 20% for all PMTs.

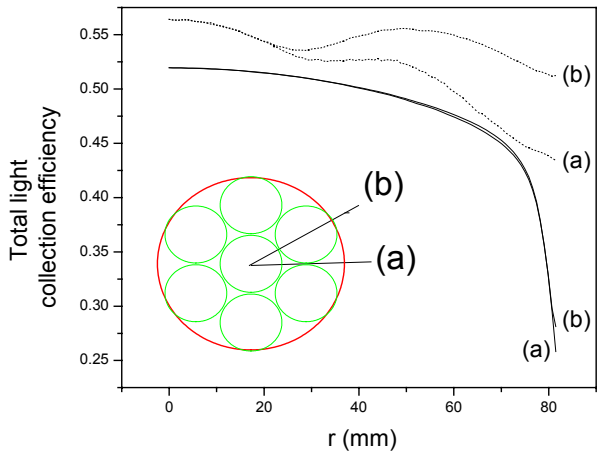

Figure 5. Total light collection efficiency, obtained by Monte Carlo simulation, as a function of the distance to the chamber axis along radial paths (a) and (b), as indicated in the insert. Solid lines are for z = 0 and dashed lines are for z = 39 mm (see Fig. 1). For z = 0 the light collection efficiencies for the two paths practically coincide.

*C. Time Resolution*

The time resolution was measured with 511 keV γ-rays with the chamber in coincidence with the $BaF_2$ detector, which gives the START signal for the TDC. The STOP signal is provided by the trigger signal of the chamber. The time resolution was determined as a function of the energy deposited by Compton scattering in the active volume. The results are shown in Table 1.

TABLE I
TIME RESOLUTION (FWHM) AS A FUNCTION OF DEPOSITED ENERGY

| Energy range (keV) | Time resolution (ns) |
|---|---|
| 70 to 105 | 3.0 |
| 50 to 70 | 3.6 |
| 35 to 50 | 4.2 |
| 20 to 35 | 4.4 |

*D. Monte Carlo*

Two independent Monte Carlo simulation models were developed. The first model is a detailed Geant4 [7] based Monte Carlo simulation of the detector and of the experimental setup. It covers several aspects of the experiment, from the generation of γ-rays in the $^{57}Co$ source and their interactions with the chamber to the production and propagation of the resulting scintillation photons in liquid xenon. For the treatment of γ-ray interactions, the low-energy electromagnetic processes of Geant4 are used, which include Compton scattering, photoelectric effect and the emission of X-rays and Auger electrons from the excited atom. The propagation of the scintillation photons in liquid xenon and their interaction with the various elements of the chamber is simulated using the optical processes, also from Geant4, in particular the Rayleigh scattering, the light absorption and the refraction/reflection.

The other model is a standalone simulation of light propagation within the active volume of the detector. It takes into account the Rayleigh scattering and the light absorption in the liquid, as well as the refraction and reflection at the optical surfaces, too, but uses independent (not from GEANT4) subroutines for those processes.

Both simulations are performed with the mean energy required to produce a scintillation photon being $W_s$ = 20 eV [4]-[6] and the quantum efficiencies of all PMTs set to 20%. The PTFE reflectivity and the absorption and scattering lengths, $\lambda_{abs}$ and $\lambda_{scat}$, respectively, for the VUV photons have been varied.

The absorption and scattering lengths for the VUV photons in liquid xenon are not known, at present. However, several authors have reported a value of about 30 cm to 40 cm for the total attenuation length (see [8], [9] and references therein). Using $\lambda_{tot}$=35 cm, which sets the constraint $1/\lambda_{tot}=1/\lambda_{scat}+1/\lambda_{abs}$, and assuming that $\lambda_{tot}=\lambda_{scat}$ (i.e. no absorption), we adjusted the simulated total number of photoelectrons to that obtained experimentally and arrive at a lower limit for the PTFE reflectivity of 0.87±0.03. This value is compatible with results reported in [10].

In a similar way, but setting the PTFE reflectivity to 1 and $\lambda_{tot}$=35 cm, we tried to estimate limits for $\lambda_{abs}$ and $\lambda_{scat}$, which would lead the simulation to agreed with experimental data. The result is a lower limit for the absorption length of about 40 cm and an upper limit for the scattering length of about 200 cm. The important point is that agreement can only be reached assuming the existence of Rayleigh scattering in liquid xenon for its scintillation light.

The above estimates were confirmed using a standalone Monte Carlo model. In this case, the ratio of the simulated total number of photoelectrons at two positions of the $^{57}Co$ source (see Fig. 5) was adjusted to that observed experimentally. The advantage of such approach is that this ratio is independent of the values set for the quantum efficiency and $W_s$, as the number of photoelectrons at each source position is proportional to their product. The result for a lower limit for the PTFE reflectivity and the limits for the absorption and scattering lengths are in excellent agreement with those obtained with the GEANT4 based model.

*E. Position Reconstruction*

A simple algorithm of 2-D position reconstruction was implemented. It is based on a Monte Carlo simulation of the light collection efficiency for each photomultiplier, for scintillation events close to the chamber bottom. The resulting amplitude distributions for the 7 PMTs are stored in a look-up table for a set of (x,y) points in steps of 1 mm. For each measured event a look-up table search was performed, using a least-squares method. Fig. 5 shows reconstructed images for 122 keV γ-rays for two positions of the source. The position resolution at the bottom center is estimated to be σ = 6.9 mm, taking into account the dimensions of the collimator.

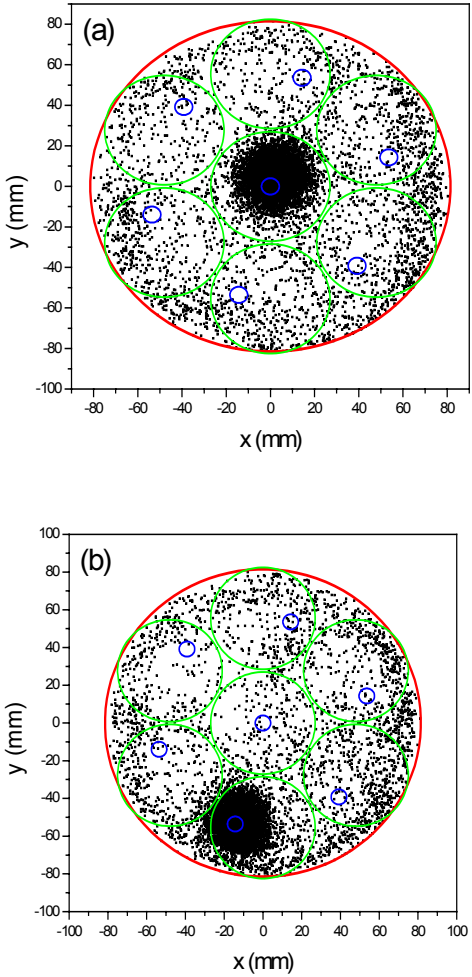

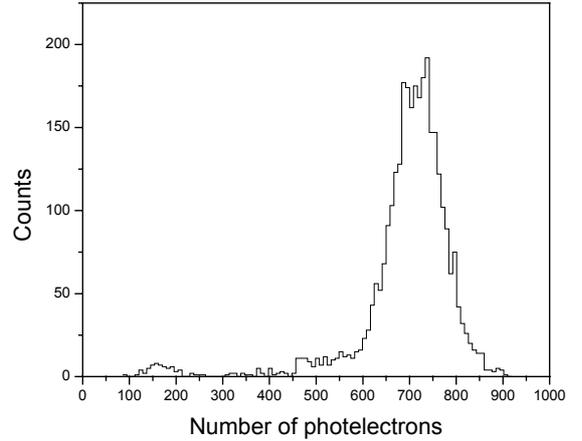

Figure 6. Reconstructed position of the scintillation points due to 122 keV γ-rays from a $^{57}$Co source, placed at the bottom of the chamber below the central collimator hole (a), and below a non-central position (b). The large circle corresponds to the active volume of liquid xenon scintillator, medium circles outline the photomultipliers and the small circles indicate the holes in the collimator, all to scale.

This algorithm was used to eliminate from the pulse height spectrum the γ-rays scattered from the chamber bottom by cutting the events with the distance from the collimator axis larger than 10 mm. The spectrum obtained is shown in Fig. 6. The comparison with Fig. 3 (a) shows a significant improvement in the quality of the spectrum thus demonstrating that those events were correctly identified and removed. Moreover, the FWHM of the 122 keV photopeak decreases to 16%. This gives us confidence on both the simulation and the analysis algorithms.

Figure 7. Pulse height spectrum of $^{57}$Co in the central position obtained by selecting events within a circle $r < 10$ mm only.

## V. CONCLUSIONS

The measurements with γ-rays show that a rather high VUV light collection efficiency is achievable with the present design, due to the PTFE surfaces surrounding the liquid xenon active volume. The total number of photoelectrons collected at the PMT photocathodes is ~5.5 photoelectrons per 1 keV of deposited energy. Energy resolutions (FWHM) of 18% and 22% were obtained with 122 keV and 511 keV γ-rays, respectively.

The time resolution was measured with 511 keV γ-rays. Values of 3.0 ns to 4.4 ns, FWHM, were obtained for deposited energies ranging from 105 keV down to 20 keV. These values are sufficient for distinguishing the fast and slow components of xenon scintillation decay.

The Monte Carlo simulation of the detector points to a PTFE reflectivity above 85% and to the existence of Rayleigh scattering of the xenon scintillation light in liquid xenon.

A position reconstruction algorithm was developed and proved to be feasible. It may be helpful also for rejecting multiple scintillation events in the active volume, for example, those due to multiple elastic scattering in experiments with neutrons. Moreover, it may lead to the improvement of the energy resolution through the correction for non-uniformity of the light collection.

The results obtained, namely the high light collection efficiency (equivalent to a low detection threshold), the time and position resolutions, together with the performance of the readout system, the thermodynamic stability and the purity of the chamber, indicate that this chamber is suitable for accurate and highly sensitive measurements with neutrons.